\documentclass[aps,prl,twocolumn,amsfonts,amssymb,amsmath,10pt,showpacs,showkeys]{revtex4-1}
\usepackage{graphicx}%
\usepackage{dcolumn}%
\usepackage{bm}%
\usepackage{bbm} 

\usepackage[usenames]{color}
\usepackage{soul} % highlights text
\usepackage[normalem]{ulem} %usage \sout{...}

\begin{document}

\title{Engineering spin waves in a high-spin ultracold Fermi gas}

\author{J.~Heinze$^{1}$}
%\affiliation{Institut f\"ur Laser-Physik, Universit\"at Hamburg, Luruper Chaussee 149, 22761 Hamburg, Germany}
%\surname{Jannes}
%\email{jheinze@physnet.uni-hamburg.de}
%\homepage{http://www.ilp.physnet.uni-hamburg.de}

\author{J.~S.~Krauser$^{1}$}
%\affiliation{Institut f\"ur Laser-Physik, Universit\"at Hamburg, Luruper Chaussee 149, 22761 Hamburg, Germany}
%\surname{Jasper}, \surname{Simon}
%\email{jkrauser@physnet.uni-hamburg.de}
%\homepage{http://www.ilp.physnet.uni-hamburg.de}

\author{N.~Fl\"aschner$^{1}$}
%\affiliation{Institut f\"ur Laser-Physik, Universit\"at Hamburg, Luruper Chaussee 149, 22761 Hamburg, Germany}
%\surname{Nick}
%\email{nflaesch@physnet.uni-hamburg.de}
%\homepage{http://www.ilp.physnet.uni-hamburg.de}

\author{K.~Sengstock$^{1,2}$}
\email[Corresponding author: ]{klaus.sengstock@physnet.uni-hamburg.de}
%\affiliation{$^{1}$Institut f\"ur Laser-Physik, Universit\"at Hamburg, Luruper Chaussee 149, 22761 Hamburg, Germany \\
%$^{2}$Zentrum f\"ur Optische Quantentechnologien, Universit\"at Hamburg, Luruper Chaussee 149, 22761 Hamburg, Germany}
%\surname{Klaus}
%\email{sengstock@physnet.uni-hamburg.de}
%\homepage{http://www.ilp.physnet.uni-hamburg.de}

\author{C.~Becker$^{1,2}$}
\affiliation{      $^{1}$Institut f\"ur Laser-Physik, Universit\"at Hamburg, Luruper Chaussee 149, 22761 Hamburg, Germany \\
			$^{2}$Zentrum f\"ur Optische Quantentechnologien, Universit\"at Hamburg, Luruper Chaussee 149, 22761 Hamburg, Germany}
%\surname{Christoph}
%\email{cbecker@physnet.uni-hamburg.de}
%\homepage{http://www.ilp.physnet.uni-hamburg.de}

\author{U.~Ebling$^{3}$}
%\affiliation{ICFO - Institut de Ci\`encies Fot\`oniques, Av.\ Carl Friedrich Gauss, 3, 08860 Castelldefels (Barcelona), Spain}
%\surname{Ulrich}
%\email{ulrich.ebling@icfo.es}
%\homepage{http://web-test.icfo.es}

\author{A.~Eckardt$^{4}$}
%\affiliation{Max-Planck-Institut f\"{u}r Physik komplexer Systeme, N\"othnitzer Str.\ 38, 01187 Dresden, Germany}
%\surname{Andre}
%\email{eckardt@pks.mpg.de}
%\homepage{http://qcqd.pks.mpg.de}

\author{M.~Lewenstein$^{3,5}$}
\affiliation{$^{3}$ICFO - Institut de Ci\`encies Fot\`oniques, Av.\ Carl Friedrich Gauss, 3, 08860 Castelldefels (Barcelona), Spain \\
$^{4}$Max-Planck-Institut f\"{u}r Physik komplexer Systeme, N\"othnitzer Str.\ 38, 01187 Dresden, Germany \\
$^{5}$ICREA-Instituci\'o Catalana de Recerca i Estudis Avan\c cats, Llu\'is Companys 23, 08010 Barcelona, Spain }

%\surname{Maciej}
%\email{maciej.lewenstein@icfo.es}
%\homepage{http://web-test.icfo.es}

\begin{abstract}

%\keywords{Quantum gases; Degenerate Fermi gases; Spin waves; High spin systems; Fundamental excitations;}
\preprint{<report number>}

We report on the detailed study of multi-component spin waves in an $s\,{=}\,3/2$ Fermi gas where the high spin leads to novel tensorial degrees of freedom compared to $s\,{=}\,1/2$ systems.
The excitations of a spin-nematic state are investigated from the linear to the nonlinear regime, where the tensorial character is particularly pronounced.
By tuning the initial state we engineer the tensorial spin-wave character, such that the magnitude and the sign of the counterflow spin currents are effectively controlled.
A comparison of our data with numerical and analytical results shows good agreement.

\end{abstract}

\maketitle

Spin-interaction driven phenomena are crucial for the behavior of many quantum systems, e.g., ferromagnets \cite{Vollhardt2001} and high-temperature superconductors \cite{Lee2006} and they are also relevant in spintronics applications \cite{Zutic2004}.
Apart from condensed matter systems with an electronic spin of $s\,{=}\,1/2$, dilute atomic gases show a wealth of novel spin excitations, where the spin is provided by the internal hyperfine structure of the atoms.
Pioneering experiments with hydrogen \cite{Johnson1984} and helium \cite{Gully1984} showed the existence of transverse spin waves, which arise from intrinsic spin-exchange interactions \cite{Bashkin1981,Lhuillier1982I,Levy1984}.
Longitudinal spin waves in two-component mixtures have been observed in non-condensed bosonic $^{87}$Rb gases \cite{McGuirk2002,Nikuni2002,Fuchs2003}.
For weakly interacting fermions, slow spin currents were reported near the zero-crossing of a Feshbach resonance \cite{Du2008,Piechon2009,Natu2009} and the interaction-induced damping of dipole oscillations was studied \cite{DeMarco2002}.
Prominent examples for spin dependent phenomena in strongly interacting fermionic systems are the miscibility of spin mixtures \cite{Sommer2011} and the quest for itinerant ferromagnetism \cite{Jo2009,Zhang2011,Conduit2011,Pekker2011}.
In contrast to conventional two-component systems, the hyperfine structure of many atoms also allows for spinor gases with $s\,{>}\,1/2$, which offer a whole new set of possibilities to study spin-dependent phenomena \cite{Ho1998,Law1998,Ohmi1998,Ho1999}. 
This includes spin-changing collisions \cite{Stenger1998,Krauser2012}, hidden interaction symmetries \cite{Wu2003,Wu2006}, spontaneous domain formation \cite{Sadler2006}, the existence of spin-nematic states \cite{Diener2006,Barnett2006}, novel superfluid phases \cite{Lecheminant2005,Rapp2007} and SU(N) degenerate ground states \cite{Honerkamp2004,Hermele2009,Cazalilla2009,Taie2010,Gorshkov2010}.
For fermionic atoms, $s\,{=}\,3/2$ constitutes the simplest realization of a high-spin system and has been thoroughly studied theoretically, being a model system for all higher spins \cite{Wu2003,Wu2006, Rodriguez2010}.

In this letter we demonstrate the controlled generation of spin waves in a quantum degenerate Fermi gas with pseudo-spin $s\,{=}\,3/2$. We experimentally study the properties of these fundamental collective spin excitations for a wide range of parameters.
The results are explained within a generalized semiclassical mean-field theory (SMFT) for fermionic atoms with a high spin of $s\,{\geq}\,3/2$, being an extension of the collisionless Boltzmann equation used to describe conventional $s\,{=}\,1/2$ systems \cite{Lhuillier1982I,Piechon2009,Natu2009,SM,Ebling2011}.
Spin waves in such high-spin systems are predicted to exhibit very complex and novel properties, which can be most intuitively understood in the language of irreducible spherical tensors (for bosonic gases, see \cite{Nikuni2008, Natu2010}).
While for $s\,{=}\,1/2$ it is sufficient to use the identity and the three spin matrices \cite{Du2008}, the description of higher spins additionally requires higher-order tensors, such as the the nematic and octupole tensor for $s\,{=}\,3/2$.

We have investigated spin-wave excitations from the linear regime, where the oscillation frequency is minimal, to the non-linear regime, where the spin-wave frequency strongly depends on the excitation amplitude.
The use of the tensor basis allows to directly observe the effect of the nonlinear mode-coupling, which leads to the excitation of breathing modes in the spin-nematic component.
Moreover, we demonstrate the controlled manipulation of the spin-wave composition by engineering the coherences of the initial state.
In that way, the spin current for two of the four components can be reversed changing the spin-wave character from spin-octupole to spin-vector. 
Our results illustrate the high degree of control that can be exerted on spin waves in high-spin Fermi gases. The good agreement with the theoretical results shows, that our SMFT well describes high-spin Fermi gases in the quantum degenerate regime.
The combined experimental and theoretical findings pave the way towards novel schemes for atom spintronics using the intrinsic high spin.
 
\begin{figure}[t]
  \centering
  \includegraphics[width=8.6cm]{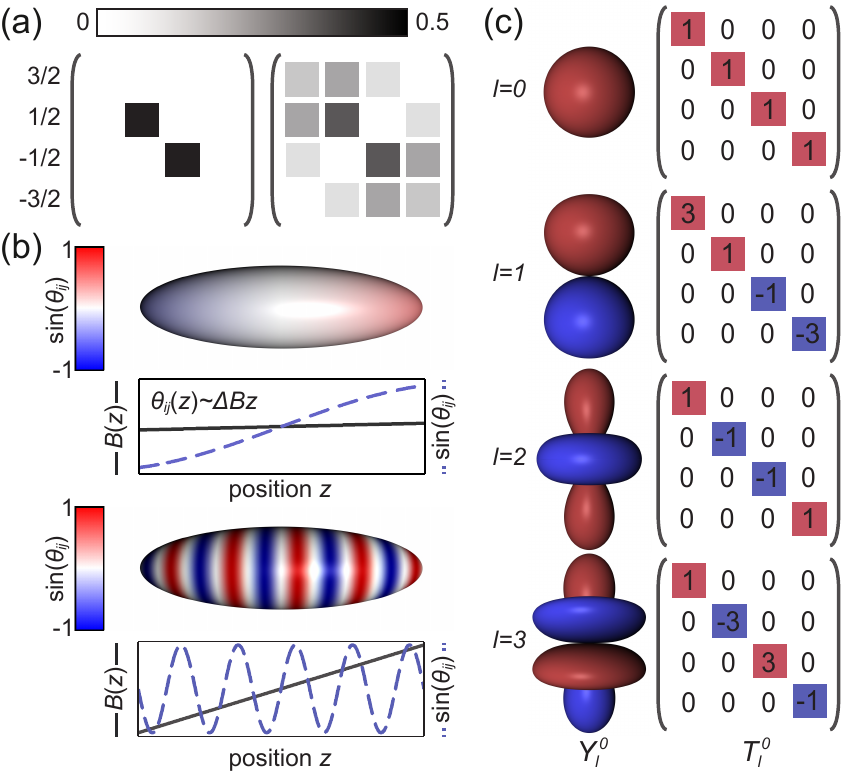}
  \caption{(a) Single-particle density matrix for an incoherent superposition of $|1/2\rangle$ and $|{-}1/2\rangle$ (left) and the resulting coherent superposition of all four components after a resonant rf-pulse (right). Diagonal elements $W_\text{ii}$ are real and represent the populations.
   Off-diagonal elements are complex numbers $W_\text{ij}\,{=}\,|W_\text{ij}|e^{i\theta_\text{ij}}$ and include the phase $\theta_\text{ij}$ between different components. Plotted is only the absolute value $|W_\text{ij}|$
   (b) Sketch of the local phase across the Fermi gas after pulses with different magnetic field gradients.
   (c) $m\,{=}\,0$ component of the $l\,{=}\,0,1,2,3$ tensor operators $T_\text{l}^\text{m}$ for $s\,{=}\,3/2$ in comparison to the corresponding spherical harmonics $Y_\text{l}^\text{m}$.}
  \label{fig1}
\end{figure}

\begin{figure}[t]
  \centering
  \includegraphics[width=8.6cm]{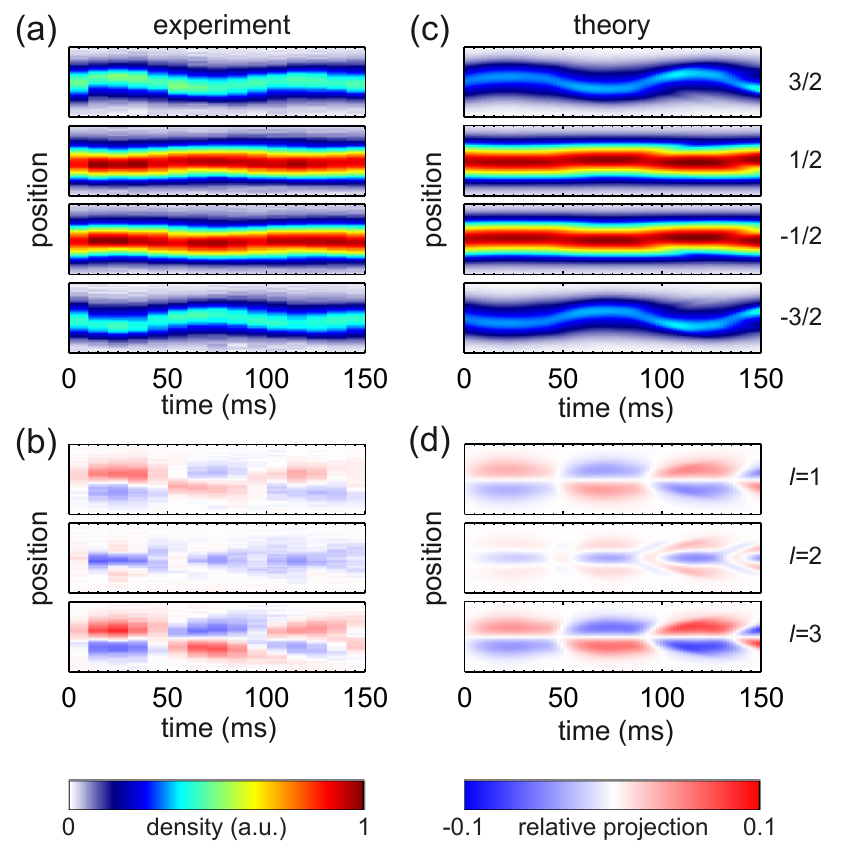}
  \caption{(a) \textit{In situ} time evolution of all four spin components after a $10\,\text{ms}$ pulse with a magnetic field gradient of $\Delta B\,{=}\,3.4\,\text{G/m}$. Shown are the column densities at different times after the excitation.
(b) Deviation from the initial population of the $m=0$ component of the vector ($l=1$), nematic ($l=2$) and octupole ($l=3$) component.
The vector and octupole component show spatial dipole oscillations, while the nematic component clearly exhibitsbreathing dynamics. 
(c,d) Numerical calculation for the parameters of (a,b).}
  \label{fig2}
\end{figure}

Our measurements are performed in a quantum degenerate gas of $^{40}$K in the $f\,{=}\,9/2$ hyperfine manifold. 
We initially evaporate a balanced mixture of $|m\,{=}\,1/2\rangle$ and $|m\,{=}\,{-}1/2\rangle$ to quantum degeneracy in an elongated, spin-independent optical dipole trap \cite{SM}. 
The final trapping frequencies are $\omega_\text{x,y,z}\,{=}\,2\pi\times(70,70,12)\, \mathrm{Hz}$. 
At this point, we apply a radio-frequency (rf) pulse to create a coherent superposition with the states $|{\pm}3/2\rangle$ [Fig.~\ref{fig1}(a)]. 
We initialize the spin waves by applying a small magnetic field gradient up to a few $\text{G/m}$ for $10\,\text{ms}$, which leads to a phase spiral for coherent superpositions of different spin components as sketched in Fig.~\ref{fig1}(b). 
While these coherent superpositions are initially still spin-polarized locally, the phase-twist allows for interactions in a trapped gas where the external potential induces spatial dynamics \cite{Ebling2011}. 
In general, the resulting mean-field interaction couples the spin degrees of freedom to different modes of the external trap leading to the emergence of spin waves. 
We detect the spin current using absorption imaging either \textit{in situ} or after $18.5\,\text{ms}$ time-of-flight (TOF) with a Stern-Gerlach separation of the spin components \cite{SM}. 
In Fig.~\ref{fig2}(a) we show a typical example for an $s\,{=}\,3/2$ spin wave initialized by a $10\,\text{ms}$ gradient of $\Delta B\,{=}\,3.4\,\text{G/m}$. 
The measurements reveal oscillatory spin currents in all four spin components.
We observe a time-independent total density, meaning that the spin waves constitute counterflow spin currents without an accompanying net mass transport.
In particular, note the inverted flow direction of $|1/2\rangle$ ($|{-}1/2\rangle$) with respect to $|3/2\rangle$ ($|{-}3/2\rangle$), which is a clear indication of the new tensorial degrees of freedom as discussed later.

For the theoretical description of high-spin Fermi gases, we generalize a 1D SMFT \cite{SM} previously used to explain spin-wave phenomena in thermal fermionic and bosonic systems with effective spin $1/2$ \cite{Lhuillier1982I,Nikuni2002,Fuchs2003,Piechon2009,Natu2009,Ebling2011} and to predict the spin-wave dynamics in thermal bosonic $s\,{=}\,1$ gases \cite{Nikuni2008, Natu2010}.
The multi-component system is described in a mean-field fashion by a single-particle density matrix (SPDM) in the form of a 1D Wigner function $W_\text{kl}(z,p)$ with spin indices $k$ and $l$. The semiclassical equations of motion take the form of a Boltzmann-equation in the collisionless regime. To leading order they read
\begin{equation}
\partial_t W(z,p) = \partial_0 W(z,p) + \frac{1}{i\hbar} [W(z,p),V(z)]\,,
\label{eq:p1}
\end{equation}
assuming a spin-independent external harmonic trap.
Here, $V_\text{mn}(z)\,{=}\,\int\sum_\text{kl} (U_\text{klnm}-U_\text{kmnl})W_\text{kl}(z,p)\,dp$ is the effective mean-field potential with the spin-dependent coupling constants $U_\text{ijkl}$ \cite{SM}, $\partial_0=(-p/m \partial_z + m\omega_z^2 z \partial_p)$ captures the time evolution due to the harmonic trap and the kinetic energy, $m$ is the mass of $^{40}$K, and $[.,.]$ indicates the commutator in spin space.
In the simulations, also higher-order terms of the mean-field interactions are taken into account, leading to very small deviations only \cite{SM}.
For an $s\,{=}\,3/2$ system, the Wigner function in spin space is a $4\times4$ SPDM, where the diagonal elements $W_\text{ii}$ represent the absolute population of the spin components and the off-diagonal elements $W_\text{ij}\,{=}\,|W_\text{ij}|e^{i\theta_\text{ij}}$ represent the single-particle coherences between different components. 
To induce a time evolution of the populations $W_\text{ii}$, it is sufficient to spatially vary the phases $\theta_\text{ij}$ of the off-diagonal elements $W_\text{ij}$, since both are coupled via the commutator in (\ref{eq:p1}).
Figure \ref{fig2}(c) shows numerical results for the exact experimental parameters, which are in good agreement with the measured results. This demonstrates the capability of the SMFT to quantitatively describe interacting high-spin Fermi gases in the quantum degenerate regime.

\begin{figure}[t]
  \centering
  \includegraphics[width=8.6cm]{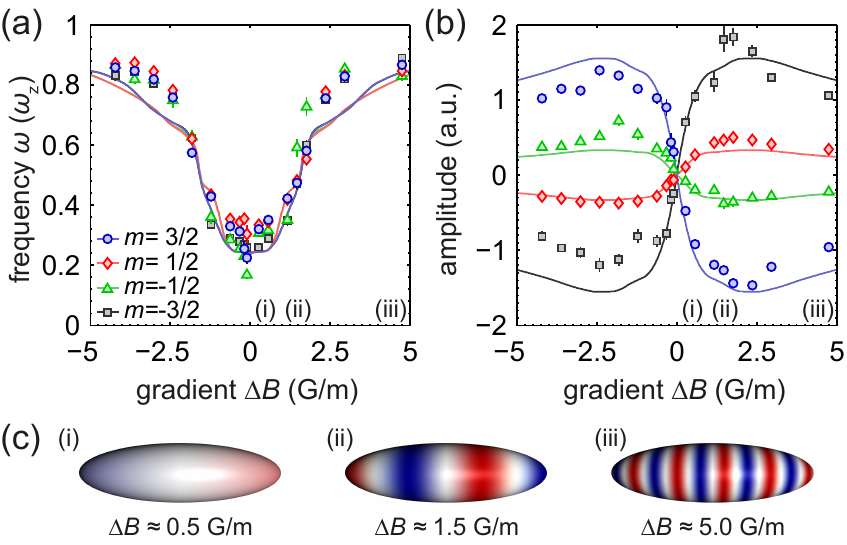}
  \caption{(a) Frequency and (b) oscillation amplitude of spin waves excited with different magnetic field gradients at $\omega_\text{z}=12\,\text{Hz}$.
  Negative amplitudes denote an inverted initial direction.
  Solid lines are numerical data for each component.
  All error bars solely correspond to fit errors, representing one standard deviation.
  The experimental amplitudes which are taken after TOF and the numerical amplitudes calculated \textit{in situ} are rescaled onto each other by a global factor.
  (c) Sketch of the phase windings across the atom cloud for different gradient pulses.} 
  \label{fig3}
\end{figure}

To obtain deeper insight into the underlying physical processes, let us at this point briefly recall the description of spin in the language of irreducible spherical tensors $T_\text{l}^\text{m}$, which simplify the equations of motion drastically.
The $T_\text{l}^\text{m}$ transform invariantly under rotations and therefore can be ordered by a total spin $l$ and a magnetic quantum number $m\,{=}\,-l,\dots,l$.
Most common are the spherical harmonics for orbital angular momentum [Fig.~\ref{fig1}(c)] and the spin-vector $\vec{S}\,{\propto}\,(\sigma_x, \sigma_y, \sigma_z)$ where $\sigma_i$ are the Pauli matrices. 
Decomposing the Wigner function (mean field) in the tensor basis as $W_\text{l}^\text{m}\,{=}\,\mathrm{Tr}(T_\text{l}^\text{m} W)$ [$V_\text{l}^\text{m}\,{=}\,\mathrm{Tr}(T_\text{l}^\text{m} V)$], the $m\,{=}\,0$ components describe the occupations whereas all other components describe coherences.
Any (pseudo) $s\,{=}\,1/2$ system can be conveniently described by the identity matrix ($l\,{=}\,0$) describing the total density, and the spin-vector $\vec{S}$ ($l\,{=}\,1$) describing the magnetization and the coherences \cite{Fuchs2003,Du2008,Natu2009,Piechon2009,Ebling2011}.
To describe the physics of larger spins it is necessary to include higher-order tensors.
In a spin $3/2$ system, as discussed here, the spin-nematic tensor ($l\,{=}\,2$) and the spin-octupole tensor ($l\,{=}\,3$) must be included (see Fig.~\ref{fig1}(c) and \cite{SM}).

Figures \ref{fig2}(b) and (d) show the time evolution of the $m\,{=}\,0$ component of the $l\,{=}\,1,2,3$ tensors for the experimental and numerical data of Fig.~\ref{fig2}(a) and (c), respectively.
Note the predominantly breathing dynamics of the spin-nematic component, qualitatively different from the spatial dipole oscillations in the spin-vector and spin-octupole component.
This results from a linear decoupling of the nematic component due to the rotational symmetry which can be understood by inserting the decomposition $W_\text{l}^\text{m}$ ($V_\text{l}^\text{m}$) into Eq.~(\ref{eq:p1}).
The rotational symmetry of the interactions leads to the particular simplification that $V_\text{l}^\text{m}\,{\propto}\,W_\text{l}^\text{m}$.
Omitting the $m$-index for simplicity the equations of motion read \cite{footnote1}:
\begin{equation}
\begin{split}
 \partial_t W_0 & \cong \partial_0 W_0\,,\\
 \partial_t W_1 & \cong \partial_0 W_1 + \frac{1}{i\hbar} \left([W_1,V_1] + [W_2,V_2] + [W_3,V_3]\right)\,,\\
 \partial_t W_2 & \cong \partial_0 W_2 + \frac{1}{i\hbar} \left([W_2,V_1+V_3] + [W_1+W_3,V_2]\right)\,,\\
 \partial_t W_3 & \cong \partial_0 W_3 + \frac{1}{i\hbar} \left([W_3,V_1] + [W_1+W_3,V_3] + [W_2,V_2]\right). 
\label{eq:p2}
\end{split}
\end{equation}
The structure of Eqs.~(\ref{eq:p2}) together with the relation $V_\text{l}^\text{m}\,{\propto}\,W_\text{l}^\text{m}$ has several important consequences. 
First, the total density $W_0^0$ is not altered by the phase spiral, since its time derivative does not depend on the off-diagonal elements; it remains constant as we observed in the experiment.
Second, the time derivative of the nematic tensors $W_2^\text{m}$ is proportional to the vector and octupole components ($W_1^\text{m}$ and $W_3^\text{m}$), but does not depend on a term $[W_2^\text{m},V_2^\text{m}]$.
This is a result of time-reversal symmetry and leads to a linear decoupling of the nematic component, in the sense that a purely nematic state does not support nematic excitations to first order.
In the nonlinear regime, however, where vector and octupole excitations possess a large amplitude, nematic excitations are created via nonlinear mode-coupling.
This leads to the weak breathing dynamics of the nematic component visible in Fig.~\ref{fig2}, where a purely nematic state was initially prepared.
The discussion above demonstrates the improved insight into high-spin spin waves granted by the irreducible spherical tensor description.

To analyze the behavior of the system for different excitation amplitudes, we applied different gradient strengths during the initialization of the spin wave \cite{McGuirk2010}.
This corresponds to a change of the initial phases $\theta_\text{ij}$ in the SPDM, while the initial coherence amplitudes $|W_\text{ij}|$ are kept constant.
In Fig.~\ref{fig3} experimental results are compared to numerical calculations and show good agreement: For small gradients, the frequency is amplitude-independent and the amplitude rises approximately linearly with the gradient strength.
For large gradients, the frequency approaches the trapping frequency and is again only weakly dependent on the excitation amplitude.
For intermediate gradients, the system shows a strongly nonlinear behavior which results in an amplitude-dependent oscillation frequency.
In the regime of small gradients, corresponding to small excitation strengths one can linearize Eqs.~(\ref{eq:p2}) and describe excitations in terms of their leading moments in $z$ and $p$ \cite{Nikuni2002}, which corresponds to pure spatial dipole oscillations \cite{SM}.
Their oscillation frequency for the present initial state can be derived to be $\omega \,{=}\, \sqrt{\omega_\text{mf}^2 + \omega_\text{z}^2}\, {-}\,\omega_\text{mf}$, where $\omega_\text{mf}$ is the mean-field interaction energy as defined in Ref.~\cite{SM}. 
This frequency is determined by a competition between the trap-induced spatial oscillations and the mean-field induced rotation of the spin.
For our parameters, we calculate $\omega=2\pi\times 2.3\,\text{Hz}$ in good agreement with the experimental results.
The linearized equations of motion also confirm the pure vector and octupole character of the dipole excitations for a perfectly nematic initial state as discussed above.
For large gradients, the frequency slowly approaches the trapping frequency: In this regime, the phase spiral is averaged out dynamically on timescales $2\pi/\omega_z$ \cite{Ebling2011} such that the mean-field potential no longer affects the subsequent oscillation.

\begin{figure}[t]
  \centering
  \includegraphics[width=8.6cm]{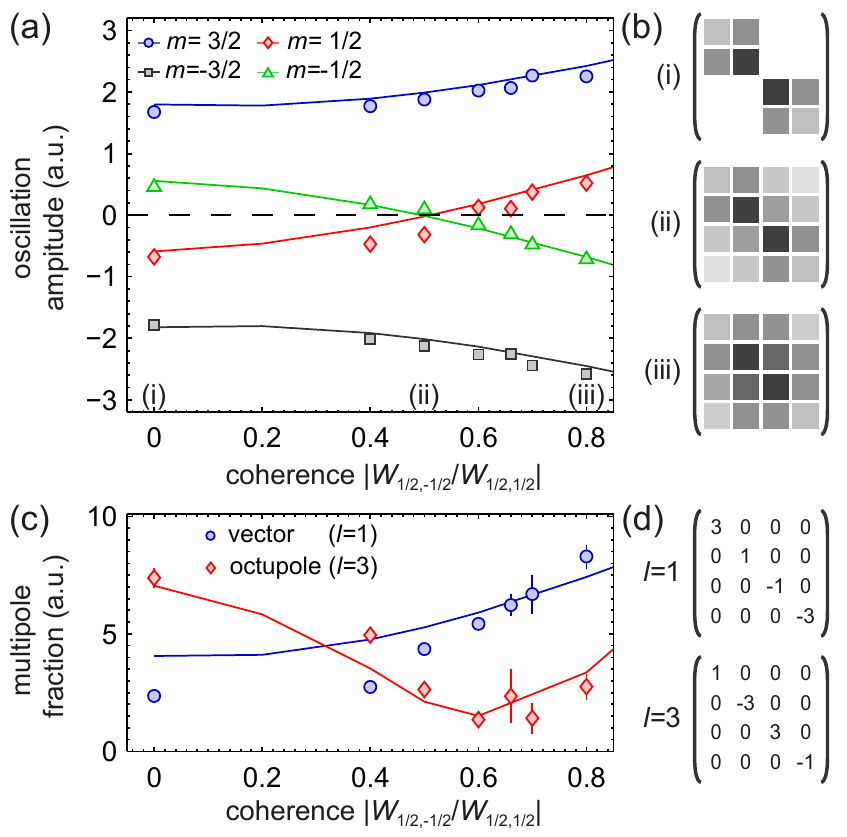}
  \caption{(a) Spatial oscillation amplitude of the spin-wave excitations for different initial coherences but equal populations of the four components at $\Delta B \,{=}\, 3.6\,\text{G/m}.$
  Solid lines show the initial spin-wave amplitude extracted from numerical calculations.
  (b) Exemplary SPDMs for different initial coherences.
  (c) Amplitude of the dipole and octupole tensor components \cite{SM}.
  (d) Vector and octupole tensors evaluated in (c).
  All error bars solely correspond to fit errors, representing one standard deviation.
  The experimental amplitudes which are taken after TOF and the numerical amplitudes calculated \textit{in situ} are rescaled onto each other by a global factor.}
  \label{fig4}
\end{figure}

All measurements discussed so far were performed with the same purely nematic initial state.
By modifying the rf-pulse sequences used for the preparation of the initial state we can control the amplitude of the coherences $|W_{ij}|$ and populations $W_\text{ii}$ in the SPDM.
By this the multipole decomposition of the initial state can be widely controlled and allows for the initialization of e.g.~pure vector or nematic initial states, which in turn results in different spin and spatial characteristics of the emerging spin wave.
Following this direction, we performed a second set of experiments, where we engineered the spin-wave excitations by keeping the population of all four spin components constant but changing the initial coherence amplitudes $|W_\text{ij}|$ [Fig.~\ref{fig4}(b)].
Note, that this is complementary to the results shown in Fig.~\ref{fig3}, where we changed the phase $\theta_\text{ij}$ of the coherence by using different gradient strengths.
Figure \ref{fig4}(a) shows the resulting oscillation amplitude for all four spin components depending on $c \,{=}\, |W_{1/2,-1/2}/W_{1/2,1/2}|$.
At $c\,{\approx}\,0.5$ the system changes its qualitative behavior where the $|{\pm}1/2\rangle$ components interchange their oscillation direction.
Using the tensor notation, the spin wave at small $c$ is dominated by the spin-octupole, where neighboring spin components have an inversed sign and therefore oscillate in opposed directions.
At large $c$, the oscillation becomes spin-vector dominated, where spin components with the same sign of magnetization oscillate in the same direction [Fig.~\ref{fig4}(c)].
The anew increase of the octupole amplitude at large $c$ is due to higher-order spatial excitations, possible in the nonlinear regime, where the measurements were performed.
At $c\,{\approx}\,0.5$, the vector and octupole component contributions mutually cancel each other, leading to a vanishing spin current in the $|{\pm}1/2\rangle$ components.
Again the numerical calculations describe the engineered spin waves very well.

In conclusion, we have thoroughly investigated the physics of collective spin waves in a high-spin Fermi gas.
Comparing experimental and numerical results, we showed that high-spin Fermi gases in the quantum degenerate regime can be well described using a SMFT.
We have analyzed the spin-wave excitation spectrum for different excitation strengths ranging from the linear deep into the nonlinear regime.
By employing irreducible spherical tensors, the SMFT allows to intuitively explain the novel emerging spin-wave characteristics in a high-spin system.
We find a linear decoupling of the spin-nematic component, which in turn allowed us to directly observe nonlinear mode-coupling in the spin-wave dynamics.
Finally, we demonstrated how to control the multipole character of spin waves which leads to a reversal of the resulting counterflow spin current of two spin components.
Our results constitute the first experimental investigation of coherent many-body dynamics of a high-spin fermionic quantum gas. They demonstrate the controlled manipulation of atomic spin currents which, together with the theoretical understanding, paves the way towards novel schemes for spintronics in ultracold atomic gases, using the intrinsic high spin as a valuable resource.
We acknowledge financial support by DFG via Grant No.~FOR801, from Spanish MICINN (FIS 2008-00784, AAII-Hubbard, FPI-fellowship), ERC grant QUAGATUA and the Spanish foundation universidad.es.

\pagebreak
\section*{Supplemental Information}

\textbf{In this supplemental material we discuss the preparation of our fermionic quantum gas (S1), the detection after TOF and \textit{in situ} (S2) and the data analysis (S3). The theoretical model is introduced in (S4) and the linearized calculation is presented in (S5). For the tensor expansion in the $s\,{=}\,3/2$ system see (S6).}

\section*{S1. Preparation of fermionic quantum gases}
We sympathetically cool about $N \,{=}\, 2 \,{\times}\, 10^{6}$ spin-polarized $^{40}$K atoms in the state $f\,{=}\,9/2$ and $|m\,{=}\,9/2\rangle$ to $0.1 \, T_{\text{F}}$ using $^{87}$Rb in a magnetic trap. Afterwards we transfer the atoms to a crossed circular-elliptical optical dipole trap with $\lambda\,{=}\,812 \, \text{nm}$.
The $1/e^2$ radii are $w_\text{x,y}\,{=}\,120 \, \mu\text{m}$ for the circular beam and $w_\text{x}\,{=}\,70\, \mu\text{m}$ and $w_\text{z}\,{=}\,280\, \mu\text{m}$ of the elliptical beam, where the tight focus is in the vertical direction.
Using rf-pulses and rf-sweeps, an equal mixture of the two hyperfine states $| 1/2\rangle$ and $|{-}1/2 \rangle$ is prepared and evaporatively cooled in the trap by a $2 \, \text{s}$ exponential intensity ramp.
The final trapping frequencies are typically $\omega_\text{x,y,z}\,{=}\,2\pi\,{\times}\,(70,70,12)\, \mathrm{Hz}$ with $N\,{=}\, 3.5\,{\times}\, 10^{5}$ particles at temperatures of $0.25 \, T_{\text{F}}$.

\section*{S2. Detection and analysis of spin components}
To obtain the experimental data, we used two different detection protocols, which have complementary advantages.
In most of the experiments, we applied time of flight imaging, where we switch off all optical potentials, and perform a Stern-Gerlach separation of the different spin states within the free expansion time of $18.5 \, \text{ms}$.
The atoms are detected via resonant absorption imaging and the center-of-mass is calculated for each of the separated clouds individually.
This method has the advantage that all spin-components can be detected simultaneously, but the disadvantage that all spatial modes but the dipole mode are effectively washed out, since the TOF mixes momentum and spatial components of these modes.

To obtain more information on the spatial modes of the excited spin-waves we employed an \textit{in situ} detection protocol, allowing us to observe, e.g., the breathing dynamics in Fig.~2(b).
Instead of separating the different spins by a Stern-Gerlach procedure in TOF, we use microwave pulses at $1.3 \, \text{GHz}$, to transfer all but one single component to the $f\,{=}\,7/2$ manifold of $^{40}$K, which is off-resonant to the detection light, and consequently the transferred atoms do not appear in the absorption images.
Afterwards we switch off all magnetic fields and optical potentials for $1\,\text{ms}$ to detect the atomic sample.
To record the time evolution of all four components, it is therefore necessary to repeat the full measurement four times.

\section*{S3. Analysis of experimental data}
For every time step of the spin-wave dynamics, we determine the center-of-mass of each of the spin components.
We extract the oscillation frequencies $\omega$ by fitting an exponentially damped cosine of the form
\begin{equation}
\text{COM}(t) =A\exp(-\Gamma t)\cos(\omega t+\Phi)+ C\,,
\end{equation}
with oscillation amplitude $A$, damping rate $\Gamma$, a phase shift $\Phi$ and a constant offset $C$.
For the dipole and octupole fraction in Fig.~4(c), we use a slightly different method.
We take the spatial average of the absolute value of the projected multipoles $M$ for each time step as
\begin{equation}
M(t) = \frac{1}{N_\text{p}} \sum_\text{p} |M(p,t)|
\end{equation}
where $p$ is the pixel index of the image and $N_\text{p}$ is the number of pixels.
By this procedure, the sign-information of the projection are lost and therefore, we fit the resulting time series by
\begin{equation}
M(t) =A\exp(-\Gamma t)\cos(\omega t+\Phi)^2+ C\,
\end{equation}
and report the amplitude $A$.

\section*{S4. Kinetic equation for a trapped high spin Fermi gas}
We describe the multi-component fermionic system using the single-particle Wigner-function, with its time evolution given by the semiclassical kinetic equation
\begin{eqnarray}
\partial_t W(z,p)=\partial_0 W(z,p)
+\frac{1}{i\hbar}\left[W(z,p), V(z)\right]\nonumber\\+\frac12\left\lbrace \partial_pW(z,p),\partial_zV(z)\right\rbrace 
\label{eq:1}
\end{eqnarray}
with the short notation $\partial_0\,{\equiv}\,{-}\,\frac{p}{m}\partial_z\,{+}\,m\omega_z^2z\partial_p$ for the single-particle part of motion. We do not explicitly include the magnetic field gradient, since it is used for the creation of the spin-wave but not necessary for its propagation once created.
The derivation of this equation is outlined in previous works \cite{Fuchs2003, Ebling2011}. One basically starts from the full mean-field dynamics of the Wigner function and applies a semiclassical approximation by neglecting terms involving higher order derivatives with respect to position and momentum. To leading (zeroth) order, the mean-field interaction gives rise to the commutator and the next (first) order is given by the anticommutator. We have neglected here even higher orders terms. The important term here is the commutator which is absent in the spinless case and describes interaction driven coherent spin dynamics. It is also dominant with respect to the anticommutator and we neglect the latter in our analytical studies (while we keep it in the numerical simulations).
The kinetic equation (\ref{eq:1}) is nonlinear, since the mean-field potential depends on the Wigner-function itself and its matrix elements are defined as
\begin{equation}
 V_{mn}(z)=\int dp \sum_{kl} (U_{klnm}-U_{kmnl})W_{kl}(z,p)
\label{eq:2}
\end{equation}
with the coupling constants defined as  $U_{ijkl}\,{=}\,\sum_{S=0,2,\ldots}^{2s-1}g_S\sum_{M=-S}^S\langle SM|ik\rangle\langle SM|jl\rangle$
for arbitrary spin $s$ with the corresponding Clebsch-Gordan coefficients $\langle SM|ik\rangle\equiv\langle SM|s,i,s,k\rangle$. Here $g_S$ denotes the interaction strength for the scattering of two particles with total spin 
$S$. For a real spin-3/2 system there are only scattering channels for $S\,{=}\,0,2$, but in the numerical simulations of (\ref{eq:1}) we take into account the values of $U_{ijkl}$ for spin 9/2 in the sub-manifold of $i,j,k,l\,{=}\,\pm 3/2,\pm1/2$, which also depend on $g_{4,6,8}$ \cite{Krauser2012}. Since in the experiment spin-waves along the $z$-axis are created, an effective one-dimensional kinetic equation is sufficient to describe these spin-waves. 
This is obtained by integrating out the transversal degrees of freedom from the three-dimensional case. The coupling constants for the 3D case,  $g^\prime_S\,{=}\,\frac{4\pi\hbar^2}{m}a_S$ with s-wave scattering lengths $a_S$, are modified by the local transversal density profile as $g_S\,{=}\,\frac{\int dx\int dy (n(x,y,z))^2}{(\int dx\int dy n(x,y,z))^2}\,{\times}\, g^\prime_S$. This modification depends on $z$ but only weakly so we approximate it with the central density $n(x,y,0)$.

\subsection{Linearized kinetic equation and moment method}
Frequencies for small amplitude spin-waves in the linear regime, such as close to the minimum in Fig.~3(a), can be obtained by linearizing the kinetic equation (\ref{eq:2}). For this, we consider small changes with respect to the stationary state $W_{mn}^0(z,p)$, namely $W_{mn}(z,p,t)\,{=}\,W_{mn}^0(z,p)+\delta W_{mn}(z,p,t)$. Thus we investigate the spin-waves for short times and small amplitudes. The mean field, as a function of density likewise expands as $V_{mn}(z,t)\,{=}\,V_{mn}^0(z)+\delta V_{mn}(z,t)$ and we substitute both expressions into (\ref{eq:1}) without the anti-commutator, to obtain the linearized kinetic equation
\begin{align}
(\partial_t-\partial_0)\delta W=&\frac{1}{i\hbar} \left(\left[\delta W,V^0\right]+\left[W^0,\delta V,\right]\right)
\label{eq:3}
\end{align}
Our next step to explicitly study different kinds of spin-wave modes is the so called \textit{moment method} \cite{Nikuni2002}. The thinking behind it is to look at the time evolution of moments of the position and momentum operators. The $l$-th moment of such an operator in the phase-space representation is the expectation value $\langle z^l\rangle_{mn}(t)\,{=}\,\int dp\int dz\delta W_{mn}(z,p,t)z^l$. Expanding the kinetic equation in moments means taking into account only different modes of spin-waves up to a certain order, e.g. dipole modes for $l\,{=}\,1$, breathing dynamics for $l\,{=}\,2$ and further.

First we need to find a suitable expression for the stationary state $W^0_{mn}(z,p)$, the state of the system before a spin-wave is excited by applying a gradient. With the preparation scheme in mind, considering the preparation pulse to be infinitely short in time, we approximate $W_0$ as a product of spin and orbital degrees of freedom $W_{mn}^0(z,p)\,{\approx}\, M_{mn} f_0(z,p)$ where $M_{mn}$ is a matrix in spin space determined by the preparation pulses and $f_0$ is the phase-space distribution for a two-component Fermi gas onto which the pulse is applied. This distribution is given for a non-interacting gas in a harmonic trap in local-density approximation as $f_0(z,p)\,{=}\,\int d^2x\,\int d^2p\,(\exp(\beta((p_x^2+p_y^2+p^2)/2m+\frac12 m (\omega_x^2 x^2+\omega_y^2 y^2+\omega_z^2 z^2)-\mu))+1)^{-1}$. 
In principle, corrections to this distribution arise due to interactions but in our case the mean-field energy is small compared to kinetic and potential energy and we neglect them. The mean field $V_{mn}^0(z)\,{=}\,\tilde M_{mn} n_0(z)$ then depends on the density distribution $n_0(z)\,{=}\,\int dp f_0(z,p)$, and we introduce the short notation $\tilde M_{mn}\,{=}\,\sum_{kl}(U_{klnm}-U_{kmnl})M_{kl}$. For the frequencies of the dipole modes we expand $\delta W$ and $\delta V$ into moments of position and momentum up to first order only
\begin{align}
\label{eq:4}
\delta W_{mn}(z,p,t)&=f_0(z,p)\left(A_{mn}(t)+zB_{mn}(t)+pC_{mn}(t)\right)\\
\label{eq:5}
\delta V_{mn}(z,p,t)&=n_0(z)\left(\tilde A_{mn}(t)+z\tilde B_{mn}(t)\right)\,.
\end{align}
We take the zeroth and first moments of $z,p$ of $\delta W$ in (\ref{eq:4}) and obtain their relationship to $A,B,C$
\begin{equation}
 A_{mn}=\frac1N\langle\openone\rangle_{mn}\,,
 B_{mn}=\frac{\langle z\rangle_{mn}}{\langle z^2\rangle_0}\,,
 C_{mn}=\frac{\langle p\rangle_{mn}}{\langle p^2\rangle_0}\,,
\label{eq:6}
\end{equation}
with particle number $N\,{=}\,\int dz\int dp f_0(z,p)$, as well as $\langle z^2\rangle_0\,{=}\,\int dz\int dp z^2 f_0(z,p)$ and $\langle p^2\rangle_0\,{=}\,\int dz\int dp p^2 f_0(z,p)$. We substitute expressions (\ref{eq:6}) into the linearized equation (\ref{eq:3}), then take again the zeroth and first moments of $z,p$ to obtain three equations for the matrices $A,\, B,\, C$ respectively
\begin{align}
\label{eq:7}
\partial_t A=&\frac{I_0}{i\hbar}\left(\left[M,\tilde A\right]+\left[A,\tilde M\right]\right)\,,\\
\label{eq:8}
\partial_t B-m\omega_z^2 C=&\frac{I_1}{i\hbar}\left(\left[M,\tilde B\right]+\left[B,\tilde M\right]\right)\,,\\
\label{eq:9}
\partial_t C+\frac1m B=&\frac{I_2}{i\hbar}\left[C,\tilde M\right]\,,
\end{align}
with coefficients $I_0\,{=}\,\frac1N\int dz n_0(z)^2$, $I_1\,{=}\,\frac1{\langle z^2\rangle_0}\int dz (zn_0(z))^2$ and $I_2\,{=}\,\frac1{\langle p^2\rangle_0}\int dz\int dp\, p^2 f_0(z,p)n_0(z)$. The first equation is decoupled from the others and does not lead to spatial dynamics, so we discard it. Equations (\ref{eq:8}) and (\ref{eq:9}) describe dipole oscillations of all spin components of the Wigner function and the frequencies can be obtained by a Fourier transform $\partial_t\rightarrow -i\omega$ and solving the eigenvalue equations, similar to the procedure for a spin 1 Bose gas in Refs. \cite{Nikuni2008, Natu2010}.

\subsection{SU(N) interactions}
For the sake of simplicity we now demonstrate this for the special case of SU(N)-symmetry ($N\,{=}\,2s+1$) assuming all scattering lengths to be equal.
For the $s=3/2$ subsystem of $^{40}$K considered in the experiments, our numerical comparison shows only small differences in the spin-wave behavior between this case and the true scattering parameters.  
In the SU(4)-symmetric case with all scattering lengths equal, $g_0\,{=}\,g_2\,{=}\,\ldots\,{\equiv}\, g$, the coupling constants are of the particularly simple form $U_{ijkl}\,{=}\,\frac g2(\delta_{ij}\delta_{kl}-\delta_{il}\delta_{kj})$.
For all matrices with a tilde in equations (\ref{eq:7}), (\ref{eq:8}) and (\ref{eq:9}) we obtain the simple expression $\tilde M\,{=}\,g\left(\mathrm{Tr}(M)\openone-M\right)$.
Further, the r.~h.~s.\ of  Eqs.\ (\ref{eq:7}) and (\ref{eq:8}) are zero and the equations for $B,C$, written in matrix form reduce to
\begin{align}
 \partial_t B -m\omega_z^2 C &= 0\\
 \partial_t C + \frac1m B &= \frac{gI_2}{i\hbar}\left[M,C\right]
\end{align}
which decouple trivially. After a Fourier transform $\partial_t\rightarrow -i\omega$ we obtain the eigenvalue equation
\begin{equation}
 (\omega^2-\omega_z^2)C=2\omega\omega_\mathrm{mf}\left[M,C\right]
\label{eq:10}
\end{equation}
where we have introduced the mean-field frequency $\omega_\mathrm{mf}\,{=}\, gI_2/2\hbar$. For all (nematic) initial spin states $M$ considered here the solutions of (\ref{eq:10}) are either the trivial case $\omega\,{=}\,\pm\omega_z$ that corresponds for instance to oscillations of the entire atomic cloud in the trap or 
\begin{equation}
\omega = -\omega_\mathrm{mf}\pm\sqrt{\omega_\mathrm{mf}^2+\omega_z^2}\,.
\end{equation}
which describe the spin-wave propagation.

\section*{S5. Tensor expansion of the Wigner function}
The methods described thus far in this supplemental material are valid for any value of the spin. We now focus on the $s\,{=}\,3/2$ case in order to demonstrate the decoupling of the spin-nematic component. In previous studies of the $s\,{=}\,1/2$ case \cite{Fuchs2003, Ebling2011} the kinetic equation and Wigner-function were rewritten in terms of the Pauli matrices $W=\frac12(W_0\openone+\vec W\cdot\vec\sigma)$ which form a complete basis for Hermitian $2\times2$-matrices and allow to identify the vector-part of the mean field as an effective magnetic field. For $s\,{=}\,3/2$ we define a similar basis $T_l^m$ comprising the identity matrix $T_0^0=\frac12\openone$, the three spin operators $T_1^m=\frac{1}{\sqrt 5}S_m$, five nematic operators $T_2^m$ and seven octupole operators $T_3^m$. For the nematicity we select all traceless symmetric matrices quadratic in $\vec S$, 
\begin{subequations}
\begin{eqnarray}
T_2^0=\frac{1}{2}\left(S_z^2-\frac54\openone\right)\\
T_2^1=\frac{1}{2\sqrt 3}\left(S_x^2-S_y^2\right)\\
T_2^2=\frac{1}{2\sqrt 3}\left(S_x S_y+S_y S_x\right)\\
T_2^3=\frac{1}{2\sqrt 3}\left(S_z S_x+S_x S_z\right)\\
T_2^4=\frac{1}{2\sqrt 3}\left(S_y S_z+S_z S_y\right)
\end{eqnarray}
\end{subequations}
and for the octupole 
\begin{subequations}
\begin{eqnarray}
T_3^0=\frac{\sqrt 5}{3}\left(S_z^3-\frac{41}{20}S_z\right)\\
T_3^1=\frac{\sqrt 5}{3}\left(S_x^3-\frac{41}{20}S_x\right)\\
T_3^2=\frac{\sqrt 5}{3}\left(S_y^3-\frac{41}{20}S_y\right)\\
T_3^3=\frac1{2\sqrt 3}\left\lbrace S_x,S_y^2-S_z^2\right\rbrace\\
T_3^4=\frac1{2\sqrt 3}\left\lbrace S_y,S_z^2-S_x^2\right\rbrace\\
T_3^5=\frac1{2\sqrt 3}\left\lbrace S_z,S_x^2-S_y^2\right\rbrace\\
T_3^6=\frac1{\sqrt 3}\left(S_x S_y S_z+S_z S_y S_x\right)\,.
\end{eqnarray}
\end{subequations}
This set of 16 Hermitian matrices  forms an orthonormal basis set with respect to the trace
\begin{equation}
\text{Tr}\left(T_l^m T_{l'}^{m'}\right)=\delta_{l l'}\delta_{m m'}\,,
\label{eq:trace}
\end{equation}
so we can expand the Wigner function into
\begin{equation}
 W(z,p)=\sum_{l=0}^3\sum_{m=0}^{2l}W_l^m(z,p)T_l^m\,,
 \label{eq:W_tensor}
\end{equation}
with coefficients $W_l^m(z,p)=\text{Tr}\left(T_l^m W(z,p)\right)$. In this basis the coupling constants of the mean field (\ref{eq:2}) have the form
\begin{equation}
 U_{abcd}-U_{adcb}=\sum_{l=0}^3\sum_{m=0}^{2l}\alpha_l \left(T_l^m\right)_{ab}\left(T_l^m\right)_{cd}\,,
\end{equation}
where the coefficients $\alpha_l$ depend on the coupling constants $g_S$. In a real $s\,{=}\,3/2$ system, $\alpha_1\,{=}\,\alpha_3$, which is a result of the SO(5) symmetry of the system \cite{Wu2003}. Analogous to (\ref{eq:W_tensor}) the mean field can be expanded as
\begin{equation}
V(z)=\sum_{l=0}^3\sum_{m=0}^{2l} V_l^m(z)T_l^m\,,
\label{eq:V_tensor}
\end{equation}
and we find, that in this basis each tensor component $V_l^m(z)=\text{Tr}\left(T_l^m V(z)\right)$ of the mean field is proportional to the respective component of the Wigner function with the coefficients $\alpha_l$
\begin{equation}
V_l^m(z)=\alpha_l\int dp W_l^m(z,p)\,.
\label{eq:V_tensor_prop}
\end{equation}
We now insert equations (\ref{eq:W_tensor}), (\ref{eq:V_tensor}) and (\ref{eq:V_tensor_prop}) into the kinetic equation (\ref{eq:1}), neglecting the anticommutator for simplicity and obtain
%\begin{widetext}
\begin{equation}
\begin{split}
\left(\partial_t-\partial_0\right)\sum_{l=0}^3\sum_{m=0}^{2l} W_l^m(z,p)T_l^m=&\\
\frac{1}{i\hbar}\sum_{l',l''}\sum_{m' m''} \alpha_{l''}  \int dq W_{l''}^{m''}(z,q)& W_{l'}^{m'}(z,p) \left[T_{l'}^{m'},T_{l''}^{m''}\right]\,.
\label{eq:11}
\end{split}
\end{equation}
%\end{widetext}
The time evolution for each tensorial component $W_l^m$ of the Wigner function is obtained from (\ref{eq:11}) by taking the trace with respect to the $T_l^m$ basis as in (\ref{eq:trace}). On the right hand side we define
\begin{equation}
\Lambda_{l'l''l}^{m'm''m} = \text{Tr}\left(T_l^m \left[T_{l'}^{m'},T_{l''}^{m''}\right]\right)\,,
\label{eq:lambda}
\end{equation}
such that (\ref{eq:11}) becomes
\begin{equation}
\begin{split}
\left(\partial_t-\partial_0\right) W_l^m(z,p)=&\\
\frac{1}{i\hbar}\sum_{l',l''}\sum_{m' m''} \alpha_{l''} & \int dq W_{l''}^{m''}(z,q) W_{l'}^{m'}(z,p) \Lambda_{l'l''l}^{m'm''m}\,.
\label{eq:wlm}
\end{split}
\end{equation}
Equation (\ref{eq:lambda}) simply states the decomposition of the commutators of tensor operators in the tensor basis itself.
Apart from the trivial case that $T_0^0$ commutes with all other tensors we know that by construction $\left[T_1^i,T_1^j\right]=i\sqrt{5}\sum_k\epsilon_{ijk}T_1^k$ and thus $\Lambda_{11l}^{ijk} = i\sqrt{5} \delta_{l1}\epsilon_{ijk}$.
Formulating an explicit expression for all values of $\Lambda_{l'l''l}^{m'm''m}$ is not straightforward so we restrict ourselves to provide those values of $l,l',l''$ for which the $\Lambda_{l'l''l}^{m'm''m}$ are non-zero for some values of $m,m',m''$:
\begin{equation}
\begin{tabular}[t]{rl}
l \vline & (l', l'')\\
\hline
0 \vline & (0,0)\\
1 \vline & (1,1), (2,2), (3,3)\\
2 \vline & (1,2), (2,1), (2,3), (3,2)\\
3 \vline & (2,2), (3,3), (1,3), (3,1)
\end{tabular}
\label{eq:tab}
\end{equation}
These relations show, which tensor components $l', l''$ contribute to the time evolution of the tensor component $l$.
To illustrate the these findings in a matrix representation we introduce the symbolic notation $W_l=\text{Tr}\left(W T_l\right) T_l$ suppressing the $m$-indices, and insert the relations (\ref{eq:tab}) into the equations of motion for $W_l$.
This brings us to equation (2) of the manuscript
\begin{equation}
\begin{split}
 \partial_t W_0 & \cong \partial_0 W_0\,,\\
 \partial_t W_1 & \cong \partial_0 W_1 + \frac{1}{i\hbar} \left([W_1,V_1] + [W_2,V_2] + [W_3,V_3]\right)\,,\\
 \partial_t W_2 & \cong \partial_0 W_2 + \frac{1}{i\hbar} \left([W_2,V_1+V_3] + [W_1+W_3,V_2]\right)\,,\\
 \partial_t W_3 & \cong \partial_0 W_3 + \frac{1}{i\hbar} \left([W_3,V_1] + [W_1+W_3,V_3] + [W_2,V_2]\right), 
\end{split}
\end{equation}
where the decoupling of the nematic spin tensor observed in the experiment is evident, since no term of the form $[W_2,V_2]$ exists on the right hand side of the equation for $\partial_t W_2$.


\begin{thebibliography}{50}
\bibitem{footnote1}
The equation is depicted symbolically. For an explicit derivation see \cite{SM}.
\end{thebibliography}
\end{document}